\documentclass{article}

\usepackage{microtype}
\usepackage{appendix}
\usepackage{rotating}
\usepackage{graphicx}
\usepackage{booktabs} % for professional tables
\usepackage{float}
\usepackage{hyperref}
\hypersetup{colorlinks,allcolors=black}
\usepackage[caption=false]{subfig}
\usepackage[accepted]{icml2020}

\icmltitlerunning{An Unsupervised Machine Learning Approach to Assess the ZIP Code LevelImpact of COVID-19 in NYC}
\begin{document}

\twocolumn[
\icmltitle{An Unsupervised Machine Learning Approach to Assess the ZIP Code Level Impact of COVID-19 in NYC}



\icmlsetsymbol{equal}{*}

\begin{icmlauthorlist}
\icmlauthor{Fadoua Khmaissia}{to}
\icmlauthor{Pegah Sagheb Haghighi}{to}
\icmlauthor{Aarthe Jayaprakash}{ed0}
\icmlauthor{Zhenwei Wu}{ed1}
\icmlauthor{Sokratis Papadopoulos}{ed2}
\icmlauthor{Yuan Lai}{ed3}
\icmlauthor{Freddy  T. Nguyen}{ed3,ed4}
%\icmlauthor{Buiui Eueu}{ed}
%\icmlauthor{Aeuia Zzzz}{ed}
%\icmlauthor{Bieea C.~Yyyy}{to,goo}
%\icmlauthor{Teoau Xxxx}{ed}
%\icmlauthor{Eee Pppp}{ed}
\end{icmlauthorlist}

\icmlaffiliation{to}{Department of Computer Science and Engineering, University of Louisville, KY, USA}
\icmlaffiliation{ed0}{School of Information Studies, Syracuse University, NY, USA}
\icmlaffiliation{ed1}{John A. Paulson School of Engineering and Applied Sciences, Harvard University, MA, USA}
\icmlaffiliation{ed2}{New York University, NY, USA}
\icmlaffiliation{ed3}{Massachusetts Institute of Technology, MA, USA}
\icmlaffiliation{ed4}{Mount Sinai Hospital, NY, USA}

%\icmlcorrespondingauthor{Fadoua Khmaissia}{c.vvvvv@googol.com}
\icmlcorrespondingauthor{}{f0khma01@louisville.edu, p0sagh01@louisville.edu, Ajayapra@syr.edu, zwu2@g.harvard.edu}

\icmlkeywords{Machine Learning, ICML}

\vskip 0.3in
]



\printAffiliationsAndNotice{}  % leave blank if no need to mention equal contribution

\begin{abstract}
%New York City has been recognized as the world's epicenter of the pandemic. The goal of this study is to determine patterns of ZIP code-level increase in number of COVID-19 cases in megacities like NYC, while considering key demographic indicators, socioeconomic status, and mobility factors.
New York City has been recognized as the world's epicenter of the novel Coronavirus pandemic. To identify the key inherent factors that are highly correlated to the Increase Rate of COVID-19 new cases in NYC, we propose an unsupervised machine learning framework. 
\begin{figure*}[h!]
\vskip 0.2in
\begin{center}
\centerline{\includegraphics[width=\textwidth]{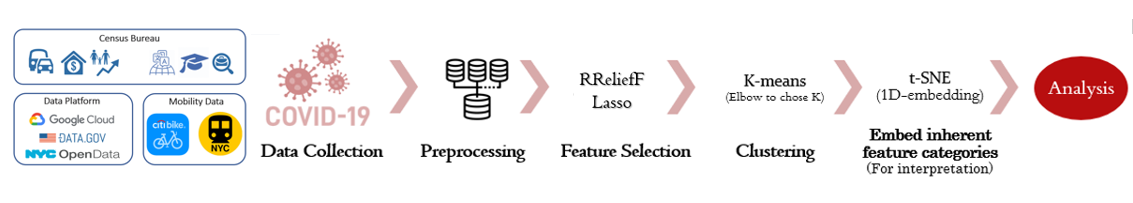}}
\caption{Proposed pipeline}
\label{pipeline}
\end{center}
\vskip -0.2in
\end{figure*}
%The goal of this study is to determine patterns of ZIP code-level increase in number of COVID-19 cases in megacities like NYC, while considering key demographic indicators, socioeconomic status, and mobility factors.
%We select the most relevant features to perform a clustering that can best reflect the spread and map them down to 9 interpretable categories.
Based on the assumption that ZIP code areas with similar demographic, socioeconomic, and mobility patterns are likely to experience similar outbreaks, we select the most relevant features to perform a clustering that can best reflect the spread, and map them down to 9 interpretable categories.
%assess the inherent risk of the contagion and inform future policies and decisions based on clusters similarities
We believe that our findings can guide policy makers to promptly anticipate and prevent the spread of the virus by taking the right measures.
% We strongly believe that being able to predictivelymap changes in the rate of infection can help pol-icy makers to promptly anticipate and prevent thespread of the virus by taking the right measures.

%Based on the assumption that regions with similar behavioral patterns are likely to experience similar outbreaks, 

%Based on the assumption that regions with similar social, demographic and mobility patterns are likely to experience similar outbreaks, our goal is to assess the intrinsic risk of the pandemic within different geo-units and inform future policies and decisions.

%The World Health Organization has declared the novel coronavirus as a global pandemic. Lock-down and  social distancing have brought urban life to a halt. New York City has been recognized as the world's epicenter of the pandemic. The goal of this paper is to determine patterns of ZIP code-level increase in number of COVID-19 cases in megacities like NYC, while considering the key demographic indicators, socioeconomic status, and mobility factors.
\end{abstract}

\section{Introduction}
\label{submission}
The World Health Organization has declared the novel Coronavirus as a global pandemic \cite{world2020coronavirus}. NYC is is one of the hardest-hit cities worldwide. Lock-down and  social distancing have brought urban life to a halt. 
%The novel coronavirus (COVID-19) has been characterized as a global pandemic by the World Health Organization .
Based on the current information provided by \cite{dowd2020demographic, garg2020hospitalization, kraemer2020effect},  demographic and health characteristics as well as mobility factors can contribute to an increased risk of getting the virus. Therefore, an early identification of the risk-prone areas based on these features is essential. In addition to understanding the link between these characteristics and COVID-19 cases, this kind of studies can be very beneficial for setting policy measures, especially in case of return-to-normal plans.

The goal of this paper is to determine patterns of ZIP code-level increase in number of new COVID-19 cases in megacities like  NYC by combining clustering and feature selection techniques.

%Based on the assumption that the population in regions having similar social,  demographic, and mobility patterns are likely to exhibit similar COVID-19 outcomes, we selected the top subset of features that can best reflect this trend and mapped them to 9 interpretable categories.

%\section{Motivations and Related Works }

%With the ongoing escalation of COVID-19, researchers throughout the world are working to better understand, mitigate, and suppress its spread. Several solutions have been introduced in order to leverage the existing potential of machine learning and tackle medical and societal challenges created by this pandemic. However, only few of them remain reliable enough to show operational impact \cite{bullock2020mapping}.  Several approaches have been proposed, ranging from sophisticated modeling to simple exploratory data analysis\cite{bullock2020mapping}. 
%So far, the key research areas have included studying the virus’s transmission, facilitating its detection, developing possible drugs and vaccines, and understanding  socio-economic impacts of the pandemic \cite{bullock2020mapping, lampos2020tracking, hu2020artificial}. 
%When it comes to assessing the societal impact, AI has been applied to answer several epidemiological research questions by modeling empirical and demographic data \cite{lampos2020tracking, hu2020artificial}. \\
%Most AI applications developed for this purpose have focused on predicting national and local statistics such as: number of new cases, mortality and recovery rate\cite{bullock2020mapping}.  

\section{Motivations}
%New York city happens to be the world’s epicenter of the COVID-19 pandemic. During these unprecedented times, the main challenge has been to better model and understand the current situation in order to inform future policies and take the right decisions. Data is one of the most important assets herein.

%Several solutions have been introduced in order to leverage the existing %potential of machine learning to investigate medical and societal challenges %created by this pandemic. Nevertheless, only few of them remain reliable %enough to show operational impact \cite{bullock2020mapping}.

Several solutions have been introduced in order to leverage the existing machine learning tools and tackle many aspects of the COVID-19 crisis \cite{bullock2020mapping}. Only few of them remain reliable enough to show operational impact \cite{bullock2020mapping}. 

In NYC, most of the existing studies were either limited to a small number of patients \cite{vaid2020machine}, \cite{richardson2020presenting}, or studying the importance of just a few factors for the outbreak \cite{hooper2020covid}, \cite{yadaw2020clinical}. 
%We chose to focus our research on NYC for multiple reasons. First,  Moreover, NYC has many novel data sources that are not currently available in other cities. This makes machine learning, and especially performing feature selection using NYC data very meaningful. 

NYC is one of the most diverse, populous, and population dense cities in the United States. Nowadays, most of the city's operations are based on data-driven governance and automated procedures. Unlike many other cities, NYC has succeeded over the last few years in making an immense quantity of its data more publicly available and actionable \cite{howard2014data}. This offers a great opportunity for the data-driven fight against the Coronavirus pandemic. For instance, performing feature selection in this context can be very useful. In fact, NYC can serve as a starting point to model and understand the spread of the pandemic. The obtained findings can be extrapolated to other regions with similar patterns.
%According to the Metropolitan Transportation Authority, around $85\%$ of the population in the United States drives to work, but in New York City, $80\%$ of the population uses public transport to commute to work \cite{authority2016mta}.

Taking these factors into consideration, we propose an unsupervised machine learning approach that, starting from a large set of features that are aggregated from different NYC datasets, aims to identify the key factors that can best model the spread of the virus at the ZIP code level. 

This research would also serve as a framework for future studies and accelerate both the geo-modeling and the forecasting of the pandemic's immanent risk.

\section{Dataset and Proposed Approach}
\subsection{Dataset}
\label{data}
Based on previous studies on factors that contribute to an increased infection risk \cite{dowd2020demographic, garg2020hospitalization, kraemer2020effect}, we quantify relevant variables from the following datasets:

The \textbf{US census data} for NYC between the year 2013-2018 at the ZIP code level which consists of 177 ZIP codes and 236 features. Some of the features in the Census data include: Basic population characteristics, household composition,  household size, etc. This dataset was retrieved from BigQuery\footnote{http://bigquery.cloud.google.com/}.
The \textbf{COVID dataset}\footnote{https://github.com/nychealth/coronavirus-data} was collected from April $4^{th}$, 2020 till May $19^{th}$, 2020. This dataset consists of the total number of positive cases per ZIP code for each day.
The \textbf{NYC subway dataset} from 2013-2018 was retrieved from BigQuery\footnote{http://bigquery.cloud.google.com/}. We calculated and included features for the total number of stations and average ridership per ZIP code.
The \textbf{Citibike dataset}\footnote{https://www.citibikenyc.com/system-data} from 2013-present includes the start station and the end station along with their respective latitude and longitudes. We calculated and included features for total number of trips to and from each ZIP code for the month of March, 2020.
The above-mentioned datasets are merged based on the ZIP code. It is worth noting that another feature called population density is also added. %as a feature to the aforementioned dataset.%
The population density is calculated by dividing using the total populations by the land area.

\subsection{Proposed Approach}
Based on the assumption that regions with similar social and demographic behaviors are more likely to exhibit similar COVID-19 outcomes, we propose to investigate how the daily increase rate of new COVID-19 cases varies across ZIP codes in NYC. 
\begin{figure*}[htb!]
\centering
%\begin{minipage}[c][1\width]{0.38\linewidth}
%\centerline{\includegraphics[width=\textwidth]{pictures/positive_case%s_overtime.png}}
%\includegraphics[width=\textwidth]{pictures/t-SNE.JPG}
%\caption{Cluster Visualization with t-SNE}
%\label{t-SNE}
%\end{minipage}
\hfill
\begin{minipage}[c][1\width]{0.38\linewidth}
\includegraphics[width=\textwidth]{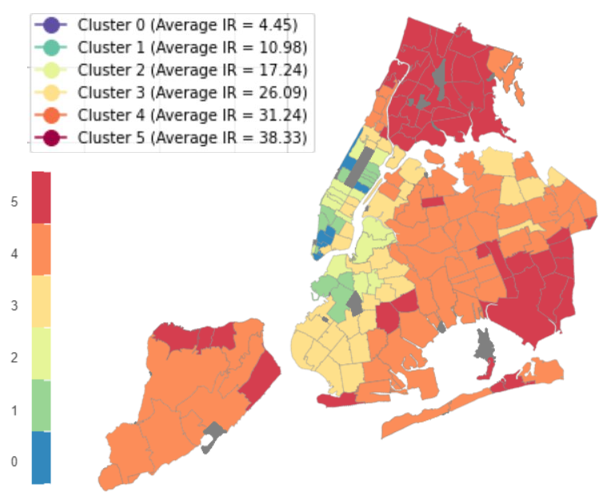}
\caption{Geo-visualization of NYC ZIP codes color-coded by Cluster ID}
\label{Clusters of ZIP codes}
\end{minipage}
\hfill
\begin{minipage}[c][1\width]{0.41\linewidth}
\includegraphics[width=\textwidth]{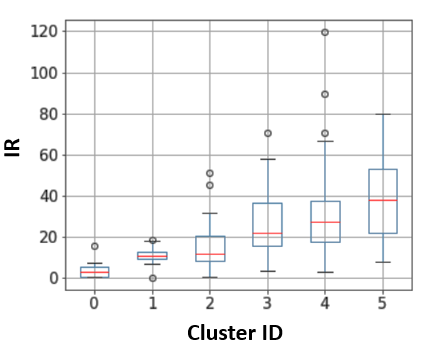}
\caption{Boxplots for the COVID-19 Increase Rates as grouped by cluster ID}
\label{Increase Rate}
\end{minipage}
\end{figure*}

\begin{figure}[t!]
%\vskip 0.15in
\begin{center}
%\centerline{
\includegraphics[width=\linewidth]{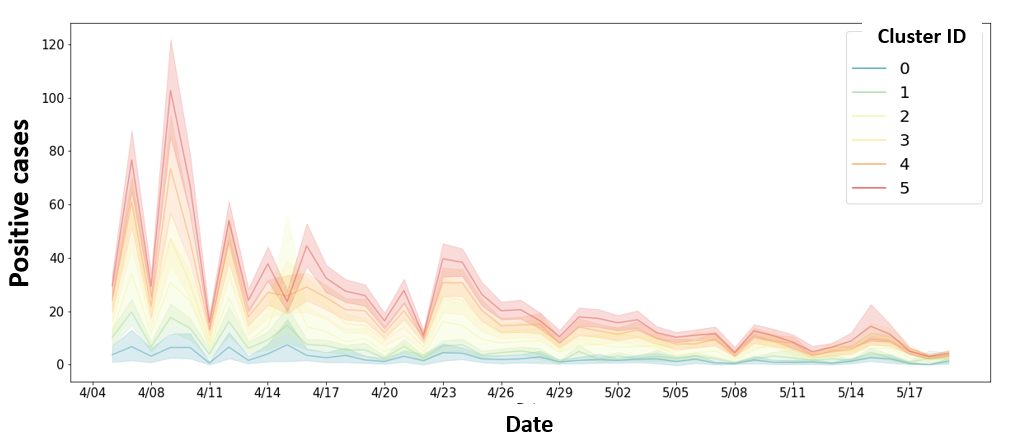}
%}
\caption{Distribution of new positive cases per cluster, overtime.}
\label{Positive cases overtime}
\end{center}
%\vskip -0.15in
\end{figure}

\begin{figure*}[h]
\vskip 0.1in
\begin{center}
\centerline{\includegraphics[width=0.91\textwidth]{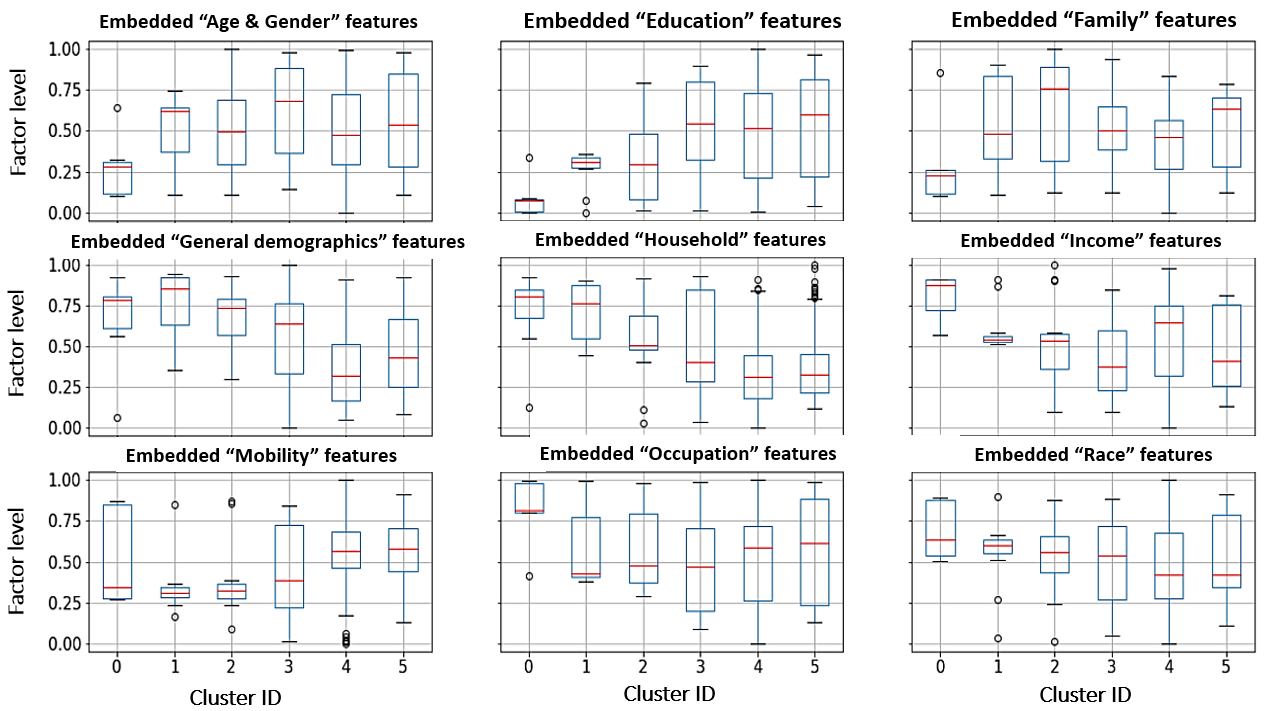}}
\caption{Boxplots for Embedded Features grouped by cluster ID}
\label{Boxplots}
\end{center}
\vskip -0.1in
\end{figure*}
We define our target as the the Average Daily Increase Rate (IR) of new COVID-19 cases. For each given ZIP code, IR is  calculated as follows:
\begin{equation}
  IR = \frac{\sum_d{NewCases(d+1) - NewCases(d)}}{N_D}  
\end{equation}
where $d \in Day_1,..,Day_{N_D}-1$ and $N_D$ is the total number of recorded days.

IR reflects the degree of acceleration of the spread of COVID-19 cases. We use it to select the most relevant features for our clustering as well as to assign the cluster IDs.
Basing our study on this target can help us assess the risk of an outbreak in a given geo-unit.

The main goal of this study is to find the optimal clusters' partition that better reflects the Increase Rate trends. The proposed  unsupervised machine learning pipeline is depicted in Figure \ref{pipeline}. 
%%%%%%
After collecting and pre-processing the data as discussed in section \ref{data}, we perform feature selection using \textbf{ Lasso technique} \cite{fonti2017feature} which, by taking into account the target variable, sets the weights of irrelevant features to zero based on $L_1$ regularization. We also perform \textbf{RReliefF} \cite{robnik1997adaptation} which estimates the strong dependencies between features and removes redundancy. The final set of features is then obtained using the union of the two selected subsets.

The optimal selected subset of features is then used for clustering. Afterwards, features within this subset are grouped into different categories (Age, Mobility, Race, etc) which will be embedded into a 1D dimension using t-SNE \cite{maaten2008visualizing} to better interpret the clusters’ behaviors.
For the reported results, we used k-means clustering  \cite{macqueen1967some}.
%which groups the data into clusters where \textit{k} is the number of clusters. 
The number of clusters $k$ was selected based on the Elbow method \cite{kodinariya2013review}.

%\begin{figure*}[h]
%\centering
%\begin{minipage}[b]{0.45\linewidth}
%\centerline{\includegraphics[width=\textwidth]{pictures/positive_cases_overtime.%png}}
%\includegraphics[width=\textwidth]{pictures/t-SNE.JPG}
%\caption{Cluster Visualization with t-SNE}
%\label{t-SNE}
%\end{minipage}
%\quad
%\begin{minipage}[b]{0.35\linewidth}
%\includegraphics[width=\textwidth]{pictures/Cluster_map_k6.png}
%\caption{Clusters of ZIP codes}
%\label{Clusters of ZIP codes}
%\end{minipage}
%\end{figure*}

\begin{table*}[ht]
\small
\caption{Factors assigned to the sub-categories of the selected feature subsets} % title of Table
\centering % used for centering table
\begin{tabular}{c|c} % centered columns (4 columns)
\hline
\bf{Mobility} & \shortstack{dwellings, no cars, more than two cars, commuters by bus,  commute $<15$ minutes, \\ commute $>15$ minutes, worked at home, station count, number of inbound trips,\\ number of outbound trips, average number riders}\\ 
\hline
\bf{Race} & {white population, black population, asian population, hispanic population, other race}\\
\hline
\bf{Education} & {associates degree, bachelors degree, high school diploma, college $<1$ year, masters degree}\\
\hline
\bf{Age and Gender} & \shortstack{male population, female population, median age, female $>50$ years old, \\ female $<50$ years old, male $>50$ years old, male $<50$ years old}\\
\hline
\bf{Family} & {single, married, divorced, widowed, parents with young children, single parent with children}\\
\hline
\bf{Income} & {median income, income $<100k$, income $>100k$, income per capita}\\
\hline
\bf{Occupation} & \shortstack{employed population, unemployed population, public administration, retail trade\\ transportation, natural resources construction, armed forces, \\ education health social, manufacturing}\\
\hline
\bf{Household} & {households, housing units, occupied housing units, vacant housing units}\\
\hline
\bf{General Demographics} & {population density}
\\[1ex] % [1ex] adds vertical space
%\hline %inserts single line
\end{tabular}
\label{table:T2} % is used to refer this table in the text
\end{table*}

%%%%

%\begin{figure*}[h!]
%%\vskip 0.2in
%%\begin{center}
%\centerline{\includegraphics[width=0.7\columnwidth]{pictures/Boxplot_IR%.png}}
%\caption{Boxplots for the Increase Rate grouped by cluster ID}
%\label{Increase Rate}
%%\end{center}
%%\vskip -0.2in
%\end{figure*}
\section{Results and Key Findings}
\subsection{Results}
% Figure \ref{Clusters of ZIP codes} shows ZIP codes with their respective clusters on the NYC map. For instance, we can see some ZIP codes with the highest number of increased rate in COVID-19 cases. One such example is Bronx having many ZIP codes colored in red. Figure \ref{Positive cases overtime} depicts the number of positive cases over time for the 6 clusters. 

After performing feature selection and combining the two subsets selected by Lasso and RReliefF, we were able to reduce the original 245 input features to just 150. Next, we run k-means clustering using $k = 6$. The cluster IDs are assigned to the different ZIP codes such that they reflect the ranking of the average IR across each cluster's ZIP codes. That is, we assign the highest cluster ID to the cluster that has the highest average IR. Figure \ref{Increase Rate} shows the distribution of the Increase Rate within each cluster. Figure \ref{Positive cases overtime} depicts how the number of positive cases has changed over time for the obtained six clusters. This plot validates our clustering results in terms of separation and compactness. In fact, the similarity of the Increase Rate between ZIP codes within the same cluster is maintained over time. Moreover, even though, the overall behavior is the same across the different clusters, the IR distributions are mostly disjoint. 
%It can be seen that there is a greater variability in terms of the Increase Rate for clusters 3, 4, and 5.

Figure \ref{Clusters of ZIP codes} illustrates the geo-visualization of our clustering results.  The map is color-coded based on the cluster IDs assigned to each of the studied NYC ZIP codes. It also reflects the geographical proximity of ZIP codes belonging to the same clusters. Bronx, Staten Island, East Brooklyn and most of Queens belong to clusters 4 and 5 which exhibit the highest IR whereas Lower Manhattan and West Brooklyn belong to Clusters 0-3 which are characterized by a slower spread. Even though our results were obtained based on an unsupervised approach, they still yield a high agreement with the statistics provided by the NYC Health Department \footnote{https://www1.nyc.gov/site/doh/covid/covid-19-data-boroughs.page} which verifies the effectiveness of our feature selection procedure.
\subsection{Key Findings}
We classify the selected subset of features into 9 categories, which we will be referring to as "factors", as summarized in Table \ref{table:T2}. In order to assess the behavior of these factors inside the different clusters, we then define 9 levels. Each factor level is computed using t-SNE 1D-embedding of all the features within the corresponding category. 

This approach aims to improve the interpretability of our findings. For instance, Figure \ref{Boxplots} displays the distribution of factor levels within the 6 obtained clusters. We notice that our factors succeed, to a large extent, in discriminating between the different clusters. 
\subsection{Discussion}
Figure \ref{Boxplots} represents the gist of our study. Keeping in mind that these embeddings do not reflect the actual weights nor the behavior of the original features, we can only use them as key indicators of which factor needs to be addressed more in order to mitigate the spread in a given geo-unit.
For instance, Clusters 4 and 5 (Bronx, some areas in Queens, East Brooklyn and Staten Island) are characterized by a relatively high level of Mobility, Education and Occupation. These zones have a high variability of both the Race and Income factors. These areas also exhibit the highest daily increase of new COVID-19 cases (Figure \ref{Increase Rate}). These trends are almost reversed for areas from Clusters 0-2.  Policy makers can refer to this embedding to gauge and eventually predict the behavior of newly affected areas. This framework can be further improved to serve as a risk assessment tool for COVID-19 outbreak.
In the appendix we provide the geo-distribution of each of the embedded feature factors across the studied NYC ZIP codes. Readers can refer to those plots for more insights.
%For instance, we notice that Cluster 0 which has the lowest IR is also exhibiting a low representation of Age,Gender and Education as opposed to Cluster 5.

%belong to clusters 4 and 5Most ZIPs in Manhatten belong to clusters 0-2 with the lowest IR.

%1. Mobility features such as commute time, dwellings, subway stops, citibike usage are very important. 2. Families with young children are more likely to get affected. 3. Male and Female above 40 are at high risk. 4. Population that are still in grades 1 to 12 or high school have a strong contribution on the Increase rate of new cases. 5. People working on transportation, sales, retails, armed forces and working parents with children are also contributing highly to the trend. 6. Poverty and low income are highly correlated to the trend. We also selected a population with income higher than 125K?!. 7. When it comes to race, asian and people who identify as other race constitute a strong factor.

% \begin{figure}[ht]
% \vskip 0.2in
% \begin{center}
% \centerline{\includegraphics[width=\columnwidth]{icml_numpapers}}
% \caption{Historical}
% \label{icml-historical}
% \end{center}
% \vskip -0.2in
% \end{figure}

%dwellings, no_cars, more than two cars, commuters by bus, commute  minutes, commute minutes, worked at home, station count, tripcount start, tripcount end, average riders
%\section{Limitations}
%As this is a work in progress, our approach still has some limitations.
Although our approach has shown promising results, it still has some limitations that need to be considered.
% - Even though the first confirmed case of COVID-19 in NYC was recorded on March $1^{st}$,  
First, our study is missing one month worth of records. The reason is that NYC Department of Health and Mental Hygiene started releasing COVID-19 statistics at a ZIP code level only from April $1^{st}$, 2020 \cite{goldstein2020coronavirus}. 
Moreover, the availability of tests in NYC was disparate and as tests became more widely accessible, the number of positive COVID-19 cases in these areas also increased.
% The inclusion of this information could potentially improve our temporal analysis by enabling us to examine different stages of the outbreak. 

% - The sample size for the American Community Survey (ACS) is smaller than that of the Decennial Census. At the ZIP code level, this sample size is even smaller. Our study assumes that the ACS Census Bureau data reliably reflects the actual demographic patterns in the city.

\section{Conclusions and Future Work}

We propose a model that combines Feature Selection and clustering techniques. Our model successfully maps similarities between ZIP codes based on mobility, socioeconomic, and demographic features with the COVID-19 daily Increase Rate trends.

%We demonstrate that by combining Feature Selection and clustering techniques we can successfully map similarities between ZIP codes based on mobility, socioeconomic, and demographic features with the COVID-19 daily Increase Rate trends.

Further work will investigate the link between IR at the ZIP code level and the availability of tests, PPE kits, and hospitals. %Examining these changes over time could assess the effectiveness of government action in the past for ZIP code level mitigation of the COVID-19 pandemic in NYC and better equip them for the future. 
Our future work will also focus on improving the model by applying other clustering techniques and by investigating more temporal patterns of the outbreak. 
Furthermore, our selected subset of features can be used to build a predictive model. 
A GUI can be designed to interactively visualize the weighted contribution of each of the input features to the embedded factors. This framework can further be extended to other regions to identify risk-prone ZIP codes.
%that need more targeted actions towards the prevention and mitigation of the pandemic.

%In addition, a prediction model based on our selected features could be developed to identify the level of risk for ZIP codes based on demographic, socio-economic, and mobility features. This model can also be extended to other megacities of the world. These methods would help identify risk-prone ZIP codes that need a more targeted approach towards the prevention and mitigation of pandemics.

\section*{Acknowledgements}
% We’d like to thank the MIT COVID-19 Challenge Datathon\footnote{https://covid19challenge.mit.edu/datathon/}, where this project began.
This work was one of the three winning projects selected  by the MIT COVID-19 Datathon committee \footnote{https://covid19challenge.mit.edu/datathon/}. 
The authors would like to express their gratitude to the Datathon organizers and to the Google team who curated multiple datasets and made them publicly available on BigQuery for this event. 
%The authors would like to acknowledge that this work was part of the MIT COVID-19 Datathon. 

\textit{The MIT COVID-19 Datathon was a week-long virtual event where teams of data scientists, clinicians, public health professionals and other subject matter experts come together to develop meaningful insights leveraging existing datasets to influence policy and decision making in the public and private sector.  }

%\textbf{Do not} include acknowledgements in the initial version of
% the paper submitted for blind review.

% If a paper is accepted, the final camera-ready version can (and
% probably should) include acknowledgements. In this case, please
% place such acknowledgements in an unnumbered section at the
% end of the paper. Typically, this will include thanks to reviewers
% who gave useful comments, to colleagues who contributed to the ideas,
% and to funding agencies and corporate sponsors that provided financial

\nocite{langley00}

\bibliography{example_paper}
\bibliographystyle{icml2020}

\newpage
%\section{Appendix}
%\providecommand{\appendixname}{Appendix}
%\appendix
\onecolumn
\begin{appendices}
\section*{Appendix}
\begin{figure*}[hbt!]
\centering
%\begin{minipage}[c][1\width]{0.38\linewidth}
%\centerline{\includegraphics[width=\textwidth]{pictures/positive_case%s_overtime.png}}
%\includegraphics[width=\textwidth]{pictures/t-SNE.JPG}
%\caption{Cluster Visualization with t-SNE}
%\label{t-SNE}
%\end{minipage}
\hfill
\begin{minipage}[c][1\width]{0.4\linewidth}
\includegraphics[width=\textwidth]{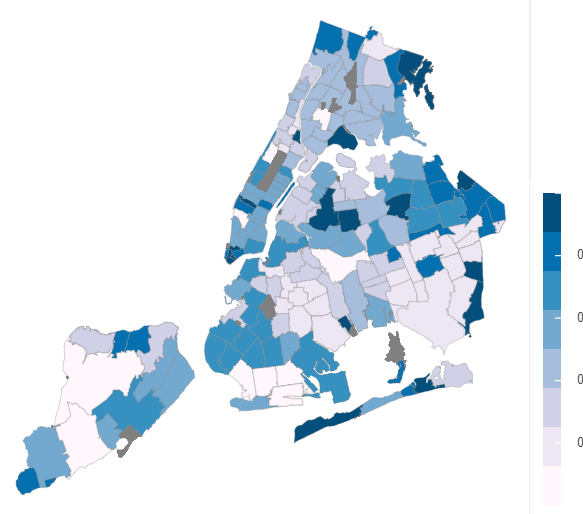}
\caption{Geo-distribution of the \textbf{Race} factor level between the different  NYC ZIP codes}
\label{Race}
\end{minipage}
\hfill
\begin{minipage}[c][1\width]{0.45\linewidth}
\includegraphics[width=\textwidth]{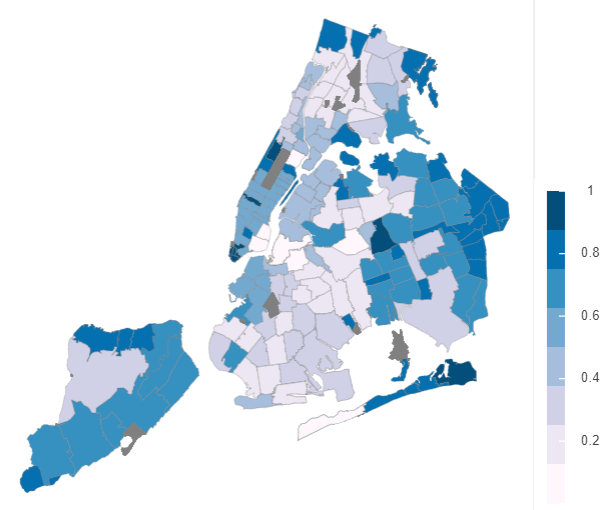}
\caption{Geo-distribution of the \textbf{Income} factor level between the different  NYC ZIP codes}
\label{Income}
\end{minipage}
\end{figure*}

\begin{figure*}[hbt!]
\centering
%\begin{minipage}[c][1\width]{0.38\linewidth}
%\centerline{\includegraphics[width=\textwidth]{pictures/positive_case%s_overtime.png}}
%\includegraphics[width=\textwidth]{pictures/t-SNE.JPG}
%\caption{Cluster Visualization with t-SNE}
%\label{t-SNE}
%\end{minipage}
\hfill
\begin{minipage}[c][1\width]{0.4\linewidth}
\includegraphics[width=\textwidth]{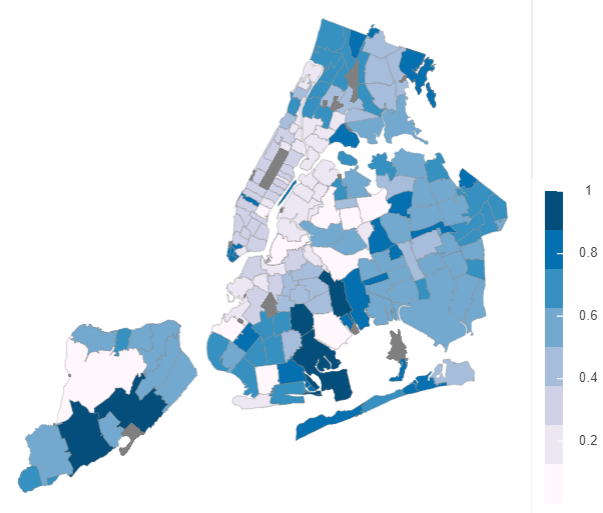}
\caption{Geo-distribution of the \textbf{Mobility} factor level between the different  NYC ZIP codes}
\label{mobility}
\end{minipage}
\hfill
\begin{minipage}[c][1\width]{0.40\linewidth}
\includegraphics[width=\textwidth]{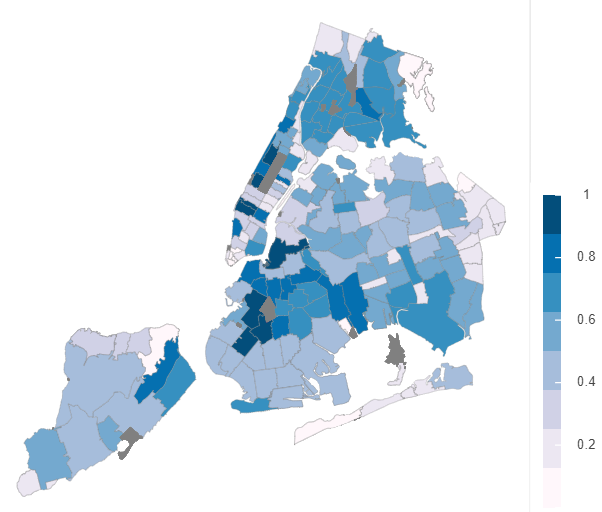}
\caption{Geo-distribution of the \textbf{Family} factor level between the different  NYC ZIP codes}
\label{Family}
\end{minipage}
\end{figure*}

\begin{figure*}[htb!]
\centering
%\begin{minipage}[c][1\width]{0.38\linewidth}
%\centerline{\includegraphics[width=\textwidth]{pictures/positive_case%s_overtime.png}}
%\includegraphics[width=\textwidth]{pictures/t-SNE.JPG}
%\caption{Cluster Visualization with t-SNE}
%\label{t-SNE}
%\end{minipage}
\hfill
\begin{minipage}[c][1\width]{0.40\linewidth}
\centering
\includegraphics[width=\textwidth]{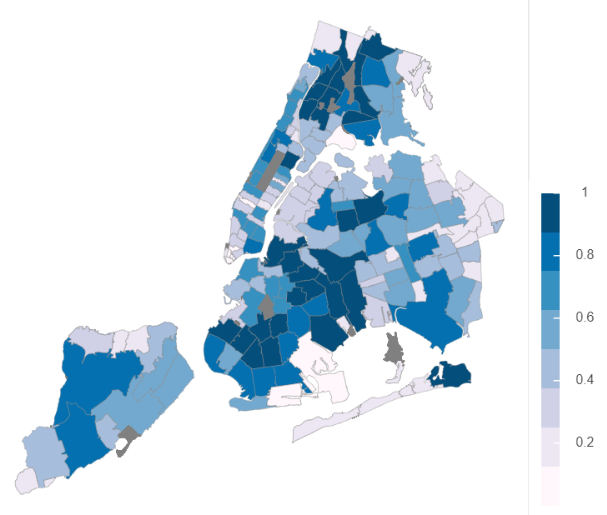}
\caption{Geo-distribution of the \textbf{Age and Gender} factor level between the different  NYC ZIP codes}
\label{Age Gender}
\end{minipage}
\hfill
\begin{minipage}[c][1\width]{0.45\linewidth}
\includegraphics[width=\textwidth]{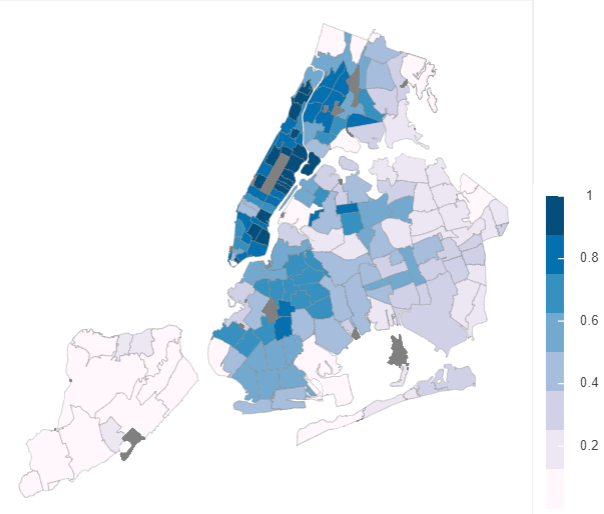}
\caption{Geo-distribution of the \textbf{General Demographics} factor level between the different  NYC ZIP codes}
\label{General Demographics}
\end{minipage}
\end{figure*}

\begin{figure*}[htb!]
\centering
%\begin{minipage}[c][1\width]{0.38\linewidth}
%\centerline{\includegraphics[width=\textwidth]{pictures/positive_case%s_overtime.png}}
%\includegraphics[width=\textwidth]{pictures/t-SNE.JPG}
%\caption{Cluster Visualization with t-SNE}
%\label{t-SNE}
%\end{minipage}
\hfill
% \begin{minipage}[c][1\width]{0.4\linewidth}
% \includegraphics[width=\textwidth]{pictures/t-SNE.JPG}
% \caption{Cluster Visualization with t-SNE}
% \label{t-SNE}
% \end{minipage}

\begin{minipage}[c][1\width]{0.45\linewidth}
\includegraphics[width=\textwidth]{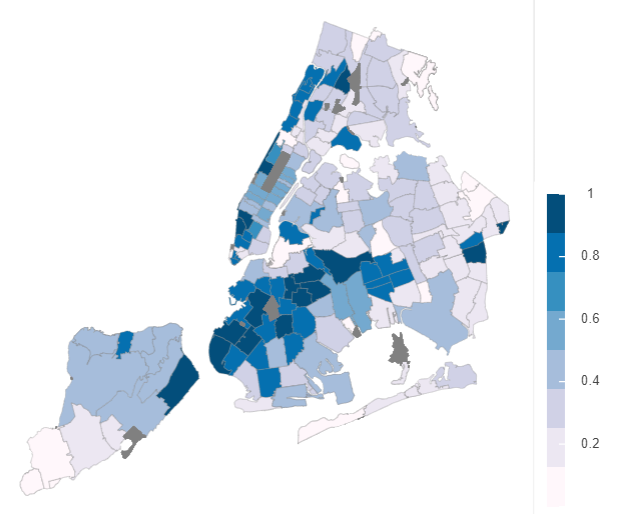}
\caption{Geo-distribution of the \textbf{Household} factor level between the different  NYC ZIP codes}
\label{Education}
\end{minipage}
\hfill
\begin{minipage}[c][1\width]{0.45\linewidth}
\includegraphics[width=\textwidth]{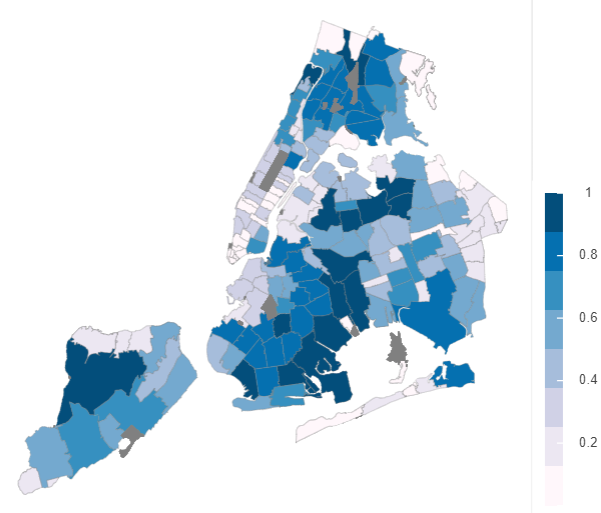}
\caption{Geo-distribution of the \textbf{Education} factor level between the different  NYC ZIP codes}
\label{Education}
\end{minipage}
\end{figure*}

\begin{figure}[htb!]
    \centering
    \begin{minipage}[c][1\width]{0.45\linewidth}
    \includegraphics[width=\textwidth]{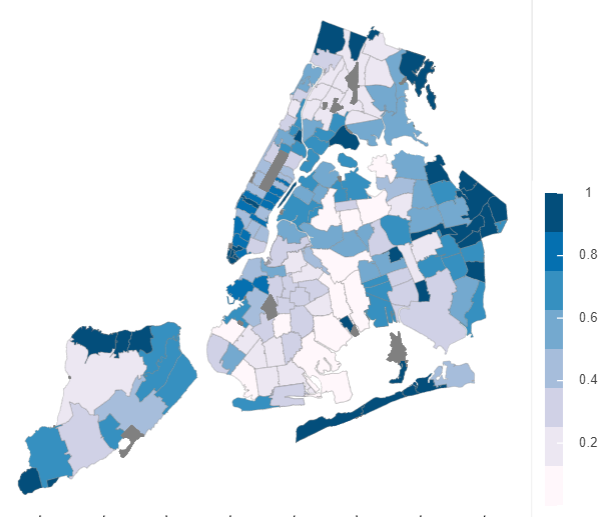}
    \caption{Geo-distribution of the \textbf{Occupation} factor level between the different  NYC ZIP codes}
    \label{Education}
    \end{minipage}

\end{figure}

\end{appendices}
\end{document}